\documentclass[conference]{IEEEtran}
\IEEEoverridecommandlockouts
\usepackage{cite}
\usepackage{amsmath,amssymb,amsfonts}
\usepackage{algorithmic}
\usepackage{graphicx}
\usepackage{epstopdf}
 \usepackage{textcomp}
\usepackage{xcolor}
\def\BibTeX{{\rm B\kern-.05em{\sc i\kern-.025em b}\kern-.08em
 T\kern-.1667em\lower.7ex\hbox{E}\kern-.125emX}}

  \makeatletter
\newcommand*{\rom}[1]{\expandafter\@slowromancap\romannumeral #1@}
\makeatother

\usepackage{xcolor}

  \usepackage[left= 1.6 cm, right= 1.6 cm, top=1.778 cm, bottom = 2.65 cm]{geometry}

\begin{document}

\title{Performance Optimization of Energy-Harvesting Underlay Cognitive Radio Networks Using Reinforcement Learning }

\author{\IEEEauthorblockN{Deemah H. Tashman\IEEEauthorrefmark{1},
Soumaya Cherkaoui\IEEEauthorrefmark{1}, and Walaa Hamouda\IEEEauthorrefmark{2}  }
\IEEEauthorblockA{\IEEEauthorrefmark{1} \small Department of Computer  and Software Engineering, Polytechnique Montreal, Montreal, Canada\\ \IEEEauthorrefmark{2}Department of  Electrical and Computer Engineering, Concordia University,  Montreal, Canada\\
Email: \IEEEauthorrefmark{1}   \{deemah.tashman, soumaya.cherkaoui\}@polymtl.ca,
\IEEEauthorrefmark{2} hamouda@ece.concordia.ca}}

\maketitle
\begin{abstract}
 In this paper, a reinforcement learning technique is employed to maximize the performance of a cognitive radio network (CRN). In the presence of   primary users (PUs), it is presumed that two secondary users (SUs) access the licensed band within underlay mode. In addition, the SU transmitter is assumed to be an energy-constrained device that requires  harvesting energy in order to transmit signals to their intended destination. Therefore, we propose that there are two main sources of energy; the interference of PUs' transmissions and ambient radio frequency (RF) sources. The SU will select whether to gather energy from PUs or only from ambient sources based on a predetermined threshold. The process of energy harvesting from the PUs' messages is accomplished via the time switching approach. In addition, based on a   deep Q-network (DQN) approach, the SU transmitter determines whether to collect energy or transmit messages during each time slot as well as  selects the suitable  transmission power in order to maximize its average data rate. Our approach outperforms a baseline strategy and converges, as shown by our findings.

\end{abstract}
\section{Introduction}

\par\IEEEPARstart{C}{ognitive}  radio networks (CRNs) have received interest as a possible solution to the issue of limited spectrum availability. CRNs function by dynamically identifying and exploiting spectrum resources, which has the potential to boost spectrum usage overall. In CRNs, there are two user groups: primary users (PUs), who have a license and have priority access to the spectrum, and secondary users (SUs), who are unlicensed and seek access to the licensed spectrum. SUs may access licensed spectrum via one of three methods: underlay, interweave, or overlay \cite{9237458}.
In underlay,  SUs operate below a certain threshold power level to avoid interfering  with the operations of the PUs. SUs transmit their signals  during the idle periods of PUs in interweave access mode. Moreover, in the overlay paradigm,  SUs can assist the PUs in their transmission in exchange for the bands' usage \cite{9926102}. The procedures executed by SUs need energy to achieve the appropriate performance level \cite{9237455}. This is a challenge, particularly for devices with limited energy resources. Consequently, SUs need to harvest energy for a variety of reasons. For instance, to boost their autonomy and decrease their reliance on an external power source. This is particularly critical for SUs operating in distant or inaccessible sites, where a steady power source may not be available. In addition, they need energy harvesting (EH) to  improve the network's sustainability. Most significantly, they must gather energy to increase the battery's life and lower the frequency with which it must be replaced. All of these factors will contribute to the enhancement of SUs' network reliability\cite{9926102}.

EH is one of the efficient strategies used to acquire energy that may be stored and utilized for several reasons, such as recharging the batteries of electronic devices, powering communications, and compensating for energy losses. Due to the simultaneous wireless information and power transmission (SWIPT) technology, research on EH-based radio frequency (RF) communications appears promising \cite{9838746}. An  RF  transmission carries both data and power, allowing the receiver to simultaneously draw power and decode messages without the need for additional equipment. To enable SWIPT, the receiver must utilize one of the EH protocols, such as time switching (TS) \cite{9612017}. This approach divides time into two distinct periods, one for energy harvesting and the other for information transfers. During the duration of information transfer, the captured energy is used to fuel the device's operations. Additionally, harvested energy may be used to charge the battery and compensating for energy losses  \cite{9500621}.

 Deep reinforcement learning (RL) blends RL with deep learning to train agents how to make choices in dynamic, complicated contexts \cite{9318243}. Agents learn from their interactions with the environment by earning rewards or penalties for their behaviors \cite{9351818}. In deep RL learning, specifically, deep neural networks are used to represent the agent's policy, value function, or environment model, enabling the agent to learn from high-dimensional and complicated observations \cite{9740504}.
 RL assists SUs in observing, learning, and taking optimum behaviors in their own local operational environment, such as dynamic channel selection and channel sensing \cite{hosey2009q}. An SU operating as an agent may optimize its spectral efficiency, energy efficiency, or throughput depending on its interaction with the surrounding environment. Deep Q-networks (DQNs) have been an intriguing strategy used by SUs in CRNs \cite{9042894}. In other words, the Q-learning algorithm's ability to converge without previous knowledge of the environment renders it well-suited for CRNs. Moreover, to improve resilience, DQNs train SUs on how to deal with unanticipated changes in the environment, such as changes in PUs' traffic or interference produced by PUs' broadcasts. This may assist in enhancing the CRN's resilience and ensuring its sustained functioning under adverse conditions \cite{hosey2009q}.

Several previous research has employed the RL approach  to enhance the reliability  of underlay CRNs. However,  a limited number of articles have suggested utilizing DQN to improve SUs' performance. For instance,  in \cite{10008522},   the authors provide an analytical model and investigation of the uplink transmission efficiency of an energy-constrained secondary sensor that operates opportunistically among several primary sensors. A deep RL-based approach, also known as a deep deterministic policy gradient (DDPG), is used to enhance the energy efficiency of the secondary sensor.
In \cite{9448276}, a deep RL approach is used to improve the transmission policies of cognitive radio-inspired non-orthogonal multiple access networks.  In conjunction with convex optimization, a DDPG is utilized to emulate the intuition of not transmitting while a PU with reliable channel broadcasts is active. Additionally, in \cite{9645987}, the authors assumed a pair of SUs and a pair of PUs, in which the SU transmitter harvests energy from ambient sources and utilized a DQN to optimize the SUs' throughput. Moreover,  the authors  examined in \cite{9680720}  an optimum transmission issue in a cognitive Internet of things  with RF EH capabilities. Proposed is a DDPG method to cope with dynamic uplink access, working mode selection, and continuous power allocation in order to optimize uplink throughput over the long term.

In contrast to the previous research, in this study,  a DQN strategy for underlay CRNs-based EH utilizing the TS protocol and ambient RF sources is utilized. In addition, this study addresses the problem of interference from the PUs and converts it into a reliable, stable supply of energy to power the transmissions of the SUs. Consequently, we suppose that an SU transmitter functioning as an agent harvests energy from two PUs and ambient RF sources, such as a cognitive base station. In addition, this SU transmitter must select whether to collect energy or broadcast messages to its intended receiver, as well as determine its transmission power. Therefore, this agent will determine, using DQNs, the best strategy that optimizes its achievable data rate. This approach can be beneficial for enhancing the efficiency and reliability of wireless communication systems, particularly in CRNs scenarios where energy constraints and spectrum sharing perform a crucial role.

 The remainder of the paper is detailed as follows; the system model is  presented in section \rom{2}. In section \rom{3}, the problem formulation is provided.  The DQN problem is expressed in section \rom{4}. Moreover, the {\color{black}simulation} results are shown in section \rom{5}. Finally,  the conclusions are provided in section \rom{6}.

\section{System Model}
\begin{figure}   [htbp]
  \centering
  \includegraphics[width=\linewidth]{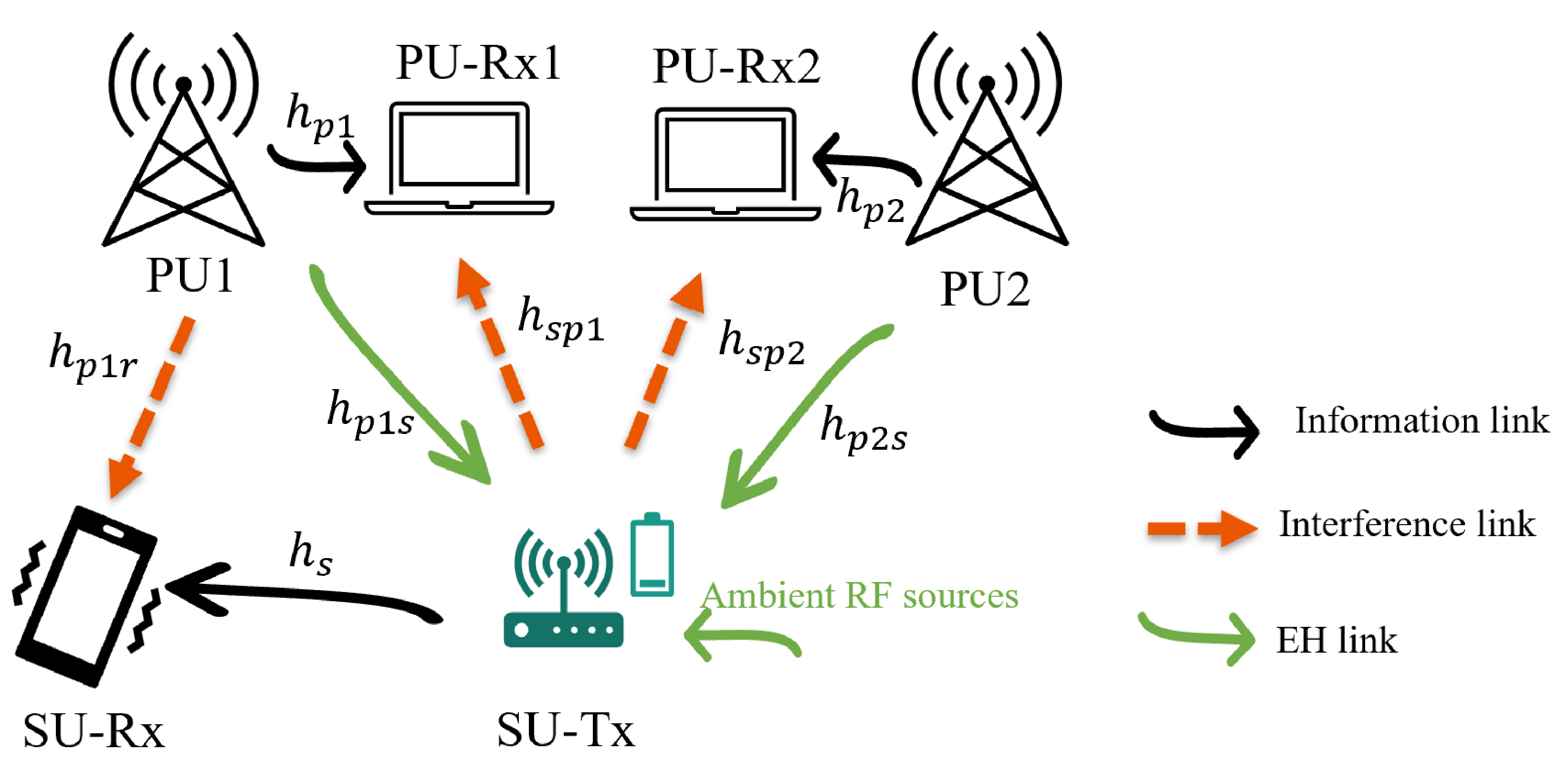}
  \caption{System model.}
  \label{sys1}
\end{figure}

As demonstrated in Fig. \ref{sys1}, an SU transmitter (SU-Tx) is communicating with an SU receiver (SU-Rx) through the underlay access mode. In this scenario, we suppose that there are two active PUs, PU1 and PU2. SU-Tx is considered to have a limited energy supply and must harvest energy to perform transmissions. Therefore, SU-Tx will collect energy from the transmissions generated by both PUs' transmitters, in addition to ambient RF sources. PU1 and PU2 are assumed to function in a slotted mode, i.e., for a total of $N$ slots, PU1 is active for the first $A$ slots, while PU2 is active for the remaining $(N-A)$ slots.  It is assumed that the SU-Tx identifies when PU1 and PU2 are active, which can be recognized by analyzing the strength of the received PUs' signals through several approaches, such as the energy detection approach \cite{9237455}. Additionally, via the transmitter verification technique \cite{4286333}, the SU-Tx can detect the position of the PUs' transmitters by analyzing the characteristics of the received signal, enabling it to differentiate between them. Moreover, SU-Tx sets a status indicator \cite{9645987}, represented by $\nu_t$ as
\begin{IEEEeqnarray}{lCr} 
\nu_t= \begin{cases}
    1, & \text{channel is occupied by PU1 in the $t^{th}$ slot} \\
    0 , & \text{channel is occupied by PU2 in the $t^{th}$ slot} 
 \end{cases} .
\end{IEEEeqnarray}
SU-Tx determines at each time slot whether to broadcast its messages or harvest energy. Consider $\kappa_t$ to be an index for the determined choice at each time and  takes  two possible values   as
\begin{IEEEeqnarray}{lCr} 
\kappa_t= \begin{cases}
    1, & \text{harvest energy in the $t^{th}$ slot} \\
    0 , & \text{transmit messages to SU-Rx in the $t^{th}$ slot} 
 \end{cases} .
\end{IEEEeqnarray}
 
\par For a more realistic scenario, the transmission power of PU1 and PU2 is considered to be uniformly distributed that ranges from 0 to $P_{max}$. The SU-Tx monitors the transmission power of PU1 and PU2 at each time slot, which can be realized through the energy detection approach \cite{9237455}. Therefore,   after detecting their transmissions power, SU-Tx determines whether to harvest RF energy from them in addition to harvesting from ambient sources, or harvesting from ambient sources only. For instance, to accomplish this, SU-Tx sets a threshold parameter $(\lambda_1)$, in which it decides to harvest energy from PU1 if the transmission power of PU1 at the $t^{th}$ time slot $(P_{p1}^t)$ was greater than $\lambda_1$. This will provide several advantages for the SUs' network. For instance, the transmission power of the PUs is regarded as interference to the communication of SUs; thus, harvesting the energy from the interference will increase the energy content of the SU-Tx battery and prevent it from PUs' interference.  In addition, if $P_{p1}^t$ is smaller than $\lambda_1$, SU-Tx will only harvest from ambient sources if it chooses to harvest and not transmit in the $t^{th}$ time slot. Featuring two energy sources ensures there is always a backup supply in case either fails. Similarly, SU-Tx establishes a threshold $(\lambda_2)$ for PU2 to operate identically to $\lambda_1$. Thus, if the transmission power of PU2 $(P_{p2}^t)$ was greater than $\lambda_{2}$, PU2 transmissions will be harvested in addition to ambient sources; otherwise, SU-Tx will only  harvest energy from ambient sources. To clarify this strategy, the EH source ($H_S$) decision is detailed below    as
\begin{IEEEeqnarray}{lCr} 
H_S= \begin{cases}
   \text {PUi + ambient sources}    , & \text{if}  P_{pi}^t \geq \lambda_i \\
    \text {Ambient sources} , &  \text{if}  P_{pi}^t < \lambda_i 
 \end{cases} , 
\end{IEEEeqnarray} 
\noindent for $i \in\{1,2\}$. Since the SUs employ the underlay access mode, the transmission power of SU-TX at each time slot $(P^t)$ should not exceed the interference threshold that the  receiver of PU1 (PU-Rx1)  and PU2 (PU-Rx2)  can tolerate.  Hence, the interference constraint can be given as \cite{9348134}
\begin{IEEEeqnarray}{lCr} 
 \nu_t g_{spi}^t P^t \leq I_{pi}^t,
\end{IEEEeqnarray}
\noindent where $I_{pi}^t$ is the interference threshold tolerable at  PU-Rx$i$, for $i \in \{1,2\}$  and $g_{spi}^t$  is the channel power gain between SU-Tx and PU-Rx$i$ and it is given as $g_{spi}^t=\left|h_{spi}^t\right|^2$, with $h_{spi}^t$ represents the channel gain  between SU-Tx and PU-Rx$i$. Moreover, let $g_m^t=\left|h_{m}^t\right|^2$, for $m=s,p1r,p1s,p2s,p1,p2,sp1,sp2$, which represents the channel power gain for the links indicated in Fig. \ref{sys1}. The  probability density function (PDF)   of  all the channels follow  the single Rayleigh model, and thereby the PDF of their channel power   gain follows the exponential distribution as 
\begin{eqnarray} \label{h22}
f_{g_m^{t}} (x)&=&   \xi_m \exp\left(-\xi_m x\right),
\end{eqnarray}
\noindent where $\xi_m$ represents the corresponding fading channel parameter. We also assume that the channel power gain  remains unchanged in each time slot for all the channel links.  When the channel is occupied by PU1, i.e., during the first $A$ time slots, the instantaneous achievable rate at SU-Rx is given as 
\begin{IEEEeqnarray}{lCr} \label{rate1}
  R_1^t=\mu T_s\log_2\left(1+\frac{P^t g_s^t}{N_0+P_{p1}^t g_{p1r}^t}\right),
\end{IEEEeqnarray}
\noindent   with $N_0$ being the variance of the additive-white-Gaussian-noise at SU-Rx.  $\mu$ value depends on the decision of SU-Tx of either harvesting from PUs or not. Thus, it is expressed  as
\begin{IEEEeqnarray}{lCr} 
\mu= \begin{cases}
    1-\rho, & \text{ harvesting from PUs and ambient   sources } \\
    1 , & \text{ harvesting  only from ambient sources} 
 \end{cases} 
\end{IEEEeqnarray}
\noindent where $0<\rho<1$ is the time switching factor of the SWIPT-EH process, in which $\rho$ of the time slot  $(T_s)$ is dedicated for EH, while the rest of time slot $((1-\rho)T_s)$ is utilized for information transmission to SU-Rx. In addition, we assume that SU-Rx is situated far away from PU2, thus its interference may be neglected at SU-Rx while PU2 is active \cite{9500621}. Therefore, {\color{black} in} the remaining slots, the rate at SU-Rx is   stated as
\begin{IEEEeqnarray}{lCr} \label{rate2}
  R_2^t=\mu T_s \log_2\left(1+\frac{P^t g_s^t}{N_0  }\right).
\end{IEEEeqnarray} 
 We assume that the maximum capacity of the battery at SU-Tx is denoted by $C_{max}$, whereas the initially available energy is given by $C_i$. The energy content of the battery will change depending on whether the SU-Tx is harvesting energy or transmitting messages, since the stored energy will be employed to power these transmissions. For example, if $\kappa_t$ equals one, SU-Tx will collect and store $E_h^t$ of energy in the battery. If SU-Tx decides to transmit, then the quantity of available energy in the subsequent time slot $(C_{t+1})$ is updated as \cite{9448276}
\begin{IEEEeqnarray}{lCr} \label{ensure}
  C_{t+1}= \min\{C_t+\kappa_t E_h^t-\left(1-\kappa_t\right) \mu P^t T_s, C_{max}\}, 
\end{IEEEeqnarray}
\noindent where $C_t$ denotes the available energy at the beginning of the $t^{th}$ slot and    $T_s$ is the duration of each time slot. $\left(1-\kappa_t\right) \mu P^t {\color{black}T_s}$ represents the utilized energy for transmission. $E_h^t=E_{TS}^t+E_{ambient}^t$, with $E_{TS}^t$ is the energy harvested from the PUs' messages using the TS protocol as
\begin{IEEEeqnarray}{lCr} 
  E_{TS}^t=\rho T_s P_{pi}^t \eta_i g_{pis}^t, 
\end{IEEEeqnarray}
\noindent for $i=\in \{1,2\}$, which corresponds to harvesting from either PU1 or PU2, respectively and      $\eta_i$ is the energy harvesting coefficient parameter. In addition, $E_{ambient}^t$ represents the energy harvested from RF ambient sources.  
Similar to \cite{9645987}, this procedure is supposed to be modeled at each time slot as a uniform distribution extending from 0 to $E_{max}$. It is worth mentioning that  (\ref{ensure}) guarantees that the stored energy will not exceed the maximum capacity of the SU-Tx battery $(C_{max})$.


\section{Problem Formulation}
 
This model's purpose is to optimise the SUs' network rate within the limits of power transmission and energy storage in the SU-Tx battery. The problem may thus be stated as follows 
  \begin{align} \label{opti-prob}
   \mathcal{P}1: \;\;  & \underset{P^t}{\text{max}}
    & & R_T \\
    & \text{s.t.} 
    & &  \label{firstrho} 0<\rho<1, &  \\
      & &&  \label{firstalpha} \sum_{t=1}^{l}P^tT_s\leq C_i+\sum_{t=0}^{l-1}E_h^t , &  \\
    & &&  \label{firstrate} (1-\kappa_t)P^t T_s\leq C_t, &  \\
    & && \label{lastcon} \nu_t g_{spi}^tP^t\leq I_{pi}^t ,
    &  
  \end{align}
\noindent where $R_T=\sum_{t=1}^{N}(1-\kappa_t)\nu_t R_1^t+(1-\kappa_t)(1-\nu_t)R_2^t$, which represents the total achievable rate at the SU-Rx in the $t^{th}$ time slot. The constraint in (\ref{firstalpha}) ensures that the total used power does not exceed the amount of energy harvested. Moreover, in (\ref{firstrate}), it is guaranteed that at the current moment, the utilized energy for transmission is less than the available energy of the battery at the beginning of the $t^{th}$ time slot. Finally, (\ref{lastcon}) ensures that the transmitted power of   SU-Tx is maintained below the interference threshold tolerable at the receivers of PU1 and PU2.   In the following section, a DQN approach is proposed to optimize this problem.
\section{DQN Approach to Optimize $P^t$}
The problem in (\ref{opti-prob})  may be represented as a model-free Markov decision process  (MDP). This is due to the fact that in each slot, the state of the current slot relies only on the state of the preceding slot, satisfying the Markov property. It is important to note that SU-Tx is unaware of the channel fading and ambient energy  exact PDF \cite{9645987}. Consequently, the model consists of the four tuples $<S, A, R, T>$. $S$ represents the states space, and thus, the state at each time slot   $S_t$  contains the following combination 
\begin{IEEEeqnarray}{lcr}\small
    S_t=\{\nu_t, E_{h_{t-1}},C_t, {\color{black} g_{s}^t}, g_{p1r}^t, g_{p1s}^t, g_{p2s}^t, g_{sp1}^t,g_{sp2}^t,g_{p1}^t,g_{p2}^t\}, \nonumber \\
\end{IEEEeqnarray}
\noindent \noindent where $E_{h_{t-1}}$ depicts the energy harvested at the previous time slot. Moreover, the action space $A$ represents the possible actions conducted by the agent. The action at each time slot  given by $A_t$ is expressed as
\begin{IEEEeqnarray}{lcr}
   A_t=\{\kappa_t, P^t\}.
\end{IEEEeqnarray}
\noindent That is, SU-Tx must select whether to collect energy or send messages, as well as  decide on the value of the transmit power $(P^t)$.  In addition, the reward function at each time slot is given by
\begin{IEEEeqnarray}{lCr} \label{reward}
\small
R_t= \begin{cases}
    R_1^t, & \kappa_t=0, \nu_t=1, P^t T_s \leq C_t, P^t g_{sp1}^t \leq I_{p1}^t  \\
    R_2^t , & \kappa_t=0, \nu_t=0, P^t T_s\leq C_t, P^t g_{sp2}^t \leq I_{p2}^t  \\
    0 , & \kappa_t=1,   P^t T_s > C_t  \\
    -\phi  , & else
 \end{cases} .
\end{IEEEeqnarray}
\noindent It is seen from 
(\ref{reward}) that multiple possibilities exist for the agent's immediate reward. The reward might be based on the achievable rates, i.e., (\ref{rate1}) and (\ref{rate2}). Moreover, the agent receives a zero reward (data rate) if it chooses to harvest energy {\color{black}and} if the transmission power exceeded the battery's available energy at the beginning of the $t^{th}$ time slot. Moreover, a penalty of $-\phi$ is incurred whenever the SU-Tx fails to adhere to any of the specified requirements. Finally, a time step $T$ is the collection of time slots. The transition from state $s_t$ to the next state $(s_{t+1})$ is termed a step. The state-action combination is repeated until all time slots are expired.

The main goal of our DQN approach is to train the agent, which is in our case the SU-Tx, and interacts with the environment to maximize the future accumulated reward \cite{9351818,9729992}. A deep neural network will estimate Q-values for each state-action combination in a given environment and approach the optimal Q-function \cite{9524882}. It learns to map state-action pairs to Q-values that represent the expected reward for taking a particular action in a given state, taking into account the future reward for following a certain policy. In DQN, the Q function can be expressed as \cite{9645987}
\begin{IEEEeqnarray}{lCr}  \label{Q-ini}
  Q^{\pi}\left(s,a\right)= E_{s^{\prime}, a^{\prime}}\left( r+\gamma Q^{\pi} \left(s^{\prime},a^{\prime}\right)|s,a\right),
\end{IEEEeqnarray}
\noindent where $s$ and $a$ are the current state and the selected action, respectively. $a^{\prime}$  is the action chosen at state $s^{\prime}$, which is the state arrived after performing the action $a$. $E[\cdot]$ is the expectation operator and $r$ represents the immediate reward for selecting action $a$. $\gamma$ is the discount factor, which   is a scalar value between 0 and 1 that determines the importance of future rewards. It determines the extent to which future rewards should be taken into account when making decisions.
Moreover, $\gamma$ is used to balance the trade-off between immediate rewards and future rewards. If $\gamma$ is close to 1, it means that future rewards are given high importance and the agent will try to maximize long-term rewards. On the other hand, if $\gamma$ is close to 0, it means that the agent only considers immediate rewards and will not favor the long-term consequences of its actions.
  The discount factor ensures that the value function converges, and the agent will always receive a finite value even when the number of time steps goes to infinity \cite{andrew1999reinforcement}. $\pi$ represents the strategy that the agent uses to select actions in a given state. The policy maps states to actions and determines which action the agent should take in each state in order to maximize the expected reward. It is worth mentioning that due to the highly dynamic
environment, the $\epsilon$-greedy strategy is
used to select the optimal transmit power of the SU-Tx. During training, the algorithm uses an $\epsilon$-greedy exploration policy, where with probability $\epsilon$ the agent selects a random action, and with probability $(1-\epsilon)$ it selects the action that maximizes the Q-value for the current state. The utilized decay equation for this policy is given as \cite{mnih2015human}
\begin{eqnarray}
    \epsilon=\epsilon_{min}+\left(\epsilon_{max}-\epsilon_{min}\right) \exp^{-d_r t},
\end{eqnarray}
\noindent where  $\epsilon_{min}$ is the minimum value of $\epsilon$, $\epsilon_{max}$ is the maximum value of $\epsilon$ (usually set to 1), and $d_r$ is the decay rate. To maximize the cumulative long-term return, we need to find the optimal action in each time slot that maximizes the state-action value defined in (\ref{Q-ini}) as
\begin{IEEEeqnarray}{lCr}  
  a^\ast =\underset{a} \arg \max  Q^{\pi}\left(s,a\right) .
\end{IEEEeqnarray}
\noindent In DQN, neural networks are employed to approximate the Q-value function  by updating the parameter $\theta$, which represents the weights and biases of the neural network. The approximation is given as 
\begin{IEEEeqnarray}{lCr}  
   Q\left(s,a;\theta \right)\approx  Q^{\ast}\left(s,a\right) ,
\end{IEEEeqnarray}
\noindent where $ Q^{\ast}\left(s,a\right)=E_{s^{\prime}} \left(r+\gamma \max_{a^{\prime}} Q^{\ast}\left(s^{\prime},a^{\prime}|s,a\right)\right)$ represents the optimal value function.

The parameter $\theta$ is updated through gradient descent and backpropagation, using the loss between the predicted Q-values and the target Q-values. The loss function $(\mathcal{L})$ is the mean square error and it is given by
\begin{IEEEeqnarray}{lCr}  \label{loss}
  \mathcal{L(\theta)}=E\left(\left( r_s+\gamma \max_{a^{\prime}} Q^{\pi} \left(s^{\prime},a^{\prime};\theta \right)-Q^{\pi}\left(s,a;\theta\right)
  \right)^2\right)   , \nonumber \\
\end{IEEEeqnarray}
\noindent where the first part of (\ref{loss})  $\left( r_s+\gamma \max_{a^{\prime}} Q^{\pi} \left(s^{\prime},a^{\prime};\theta \right)\right)$ represents the target Q-value and the second part is the predicted Q-value, which are updated periodically during the learning process. {\color{black} $r_s$ represents the reward received when moving from state $s$ to state $s^\prime$.}
Stochastic gradient descent (SGD) can be applied on the loss function to minimize it, 
 as it updates the value of $\theta$  in the direction that reduces the loss. The SGD rule is give as \begin{IEEEeqnarray}{lCr}  \label{sgd}
y=\alpha \frac{\partial  \mathcal{L(\theta)} }{\partial \theta}    ,
\end{IEEEeqnarray}
\noindent where $0<\alpha<1$ represents the learning rate, which  determines how quickly the Q-value function is updated based on the observed rewards and the current estimates of the Q-value function. {\color{black}Finally, using} (\ref{sgd}), $\theta$ is updated as  
\begin{IEEEeqnarray}{lCr}   
\theta=\theta-y .
\end{IEEEeqnarray}
 \section{{\color{black}Simulation} Results} 
\par Herein, we evaluate the proposed scenario through simulations.  Through the entire analysis, we assume that $P_{p1}^t$ and $P_{p2}^t$ are varying randomly in the range from $0$ to $1$ Watt $(P_{max})$. Moreover, we assume that $\xi_s=\xi_{p1s}=\xi_{p1r}=\xi_{p2s}=\ \xi_{p1}=\xi_{p2}=\xi_{sp1}=\xi_{sp2}=0.1$, {\color{black} $\lambda_1^t=\lambda_2^t=\lambda$}, $N=20$, $I_{P1}^t=I_{P2}=0.5$ Watt, $\gamma=0.99$, $\alpha=0.003$, $\eta=0.9$, $T_s=1$, $N_0=1$, $E_{max}=0.2$ Joule, and $\phi=1$. $\epsilon_{max}=1$, $\epsilon_{min}=0.01$, and  
 $d_r=0.001$.

\begin{figure}   [b!]
  \centering
  \includegraphics[width=0.8\linewidth]{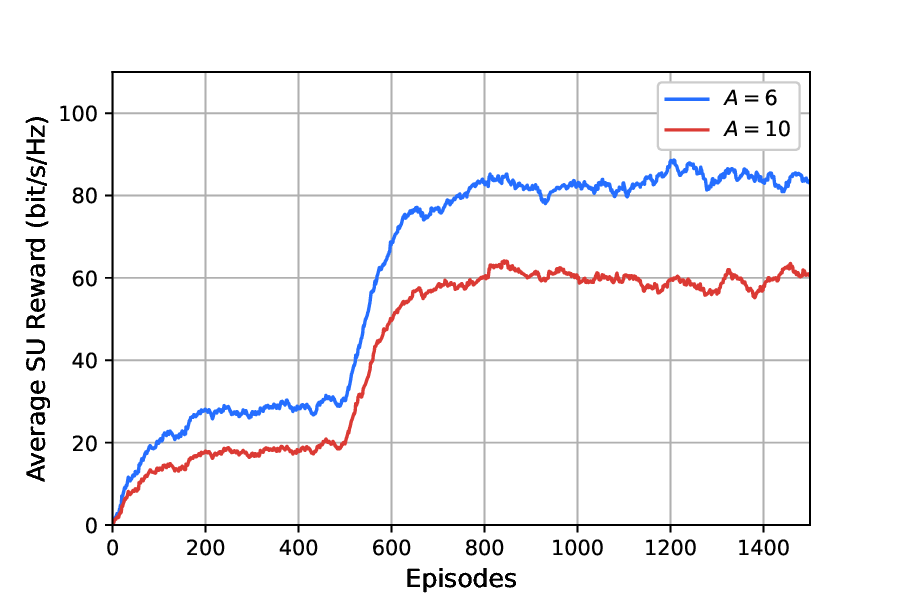}
  \caption{Average SUs' reward. $\rho=0.5$, $\lambda=0.1$,  and $C_{max}=0.5$ Joule.}
  \label{fig1}
\end{figure}

Fig. \ref{fig1} shows the average SU reward versus the number of episodes for two values of the transmission slots dedicated to PU1 $(A)$. Using the suggested DQN method, the convergence is observed, implying that the DQN optimizes long-term performance. In addition, the average reward declines as the number of slots used by the PU1 for transmissions  increases. This is because the SU receiver is impacted by the interference of PU1 and as it broadcasts for a longer duration, i.e. a larger $A$, the interference affecting SU-Rx will increase. This will lead to a reduction in the total rate and thus the average reward.

\begin{figure}   [t!]
  \centering
  \includegraphics[width=0.8\linewidth]{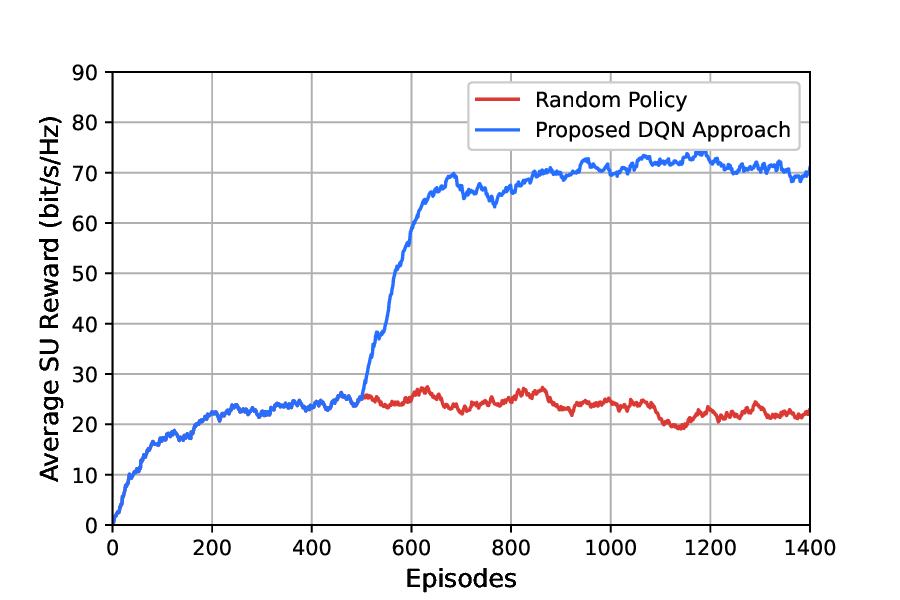}
  \caption{Average SUs' reward.  $A=10$, $\rho=0.4$,   $\lambda=0.1$, and $C_{max}=0.5$ Joule.  }
  \label{fig3}
\end{figure}

Fig. \ref{fig3} compares our proposed DQN strategy with a random policy. A random policy depicts a situation in which the agent chooses an action randomly without learning from its environment. In other words, at each time slot, the agent decides at random whether to transmit or harvest and selects a random transmission power value. During the first phase of the agent's investigation, it is noted that the results of both policies match precisely. This is due to the fact that at the beginning of the agent's training phase under the DQN strategy, it begins exploring the environment rather than exploiting the most favorable options. This is due to adopting $\epsilon$-greedy policy, in which at the beginning the agent explores with $\epsilon=1$. This means that the agent will take random actions with probability of 1, and thus explore the environment fully.  However, as the agent becomes more familiar with the environment and begins to learn which actions are more likely to lead to high rewards, a decrease in the value of $\epsilon$ begins.   Once the agent discovers the best actions, it will begin to exploit the optimal decisions, which will provide the highest future return and exceed the random strategy.   

 \par
\begin{figure}   [b!]
  \centering
  \includegraphics[width=.8\linewidth]{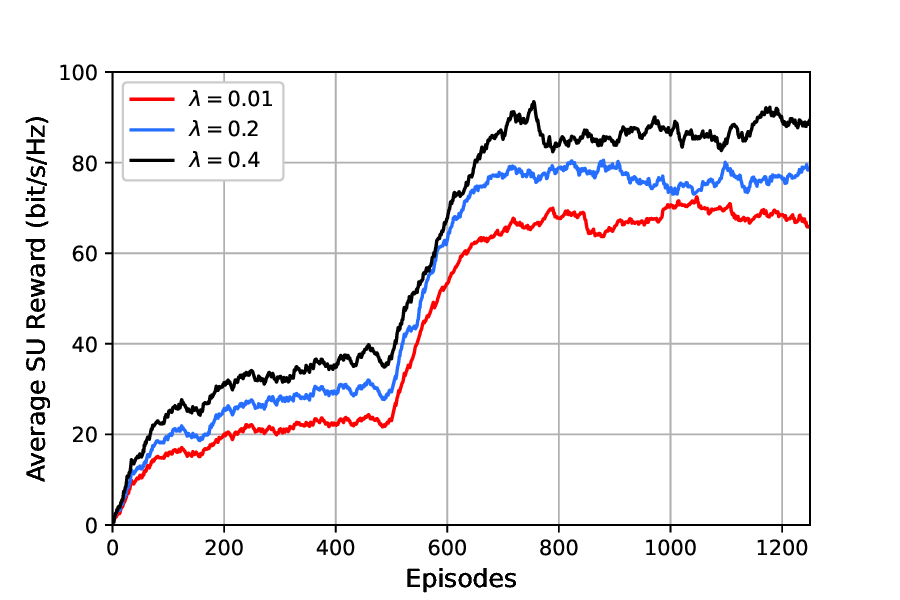}
  \caption{Average SUs' reward.  $A=10$, $\rho=0.4$,    and $C_{max}=0.5$ Joule.}
  \label{fig2}
\end{figure}

  Fig. \ref{fig2} shows the average SU rewards for various  threshold levels $(\lambda)$. It can be concluded that as $\lambda$ increases, which indicates the threshold set by SU-Tx to decide whether to harvest from PUs or not, the average SU reward improves. Recall that the SU-Tx is harvesting energy from PU transmissions through the TS protocol, and the time slot is thus split between harvesting and transmission depending on the time switching factor $\rho$. Hence, as $\lambda$ increases, less energy will be gathered through the TS protocol, while more energy will be harvested from ambient resources. Harvesting from ambient sources  will take into account the whole time slot  for the transmission, resulting in a higher attainable rate at the SU-Rx. {\color {black} Nonetheless, for a lower $\lambda$, the SU-Tx battery has more energy to account for power losses and has benefited from the interference of the PUs' broadcasts as a second energy source through TS-EH.}
   
\begin{figure}   [t!]
  \centering
  \includegraphics[width=0.8\linewidth]{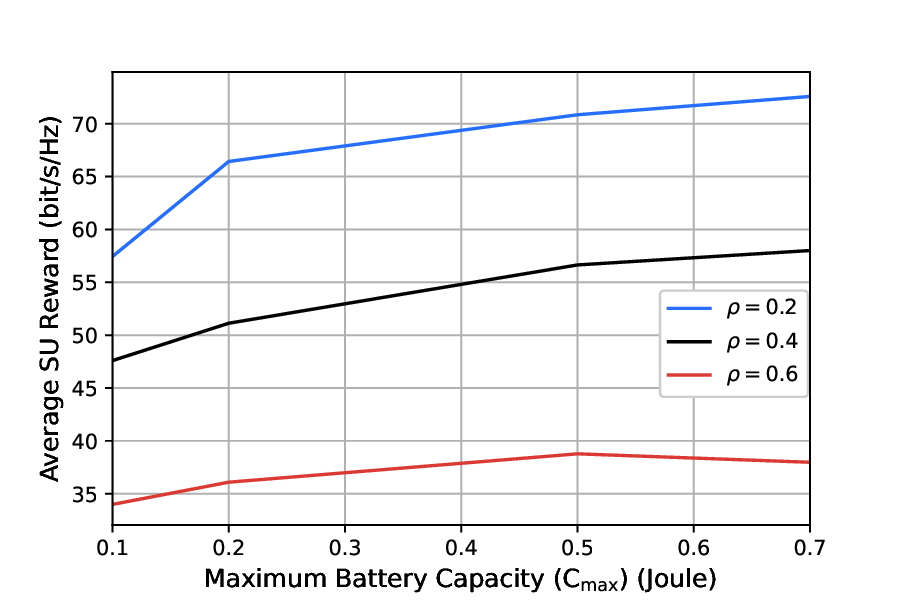}
  \caption{Average SUs' reward versus the maximum battery  capacity $C_{max}$.  $A=10$ and $\lambda=0.1$.}
  \label{fig4}
\end{figure}

Fig. \ref{fig4} illustrates how the maximum capacity of the SU-Tx battery ($C_{max}$) affects the system. It is observed that the average rate of SUs improves as $C_{max}$ rises. This is due to the fact that a larger battery capacity indicates that the SU-Tx can collect a greater quantity of energy, resulting in a higher amount of transmission power. This will increase the signal-to-noise ratio at SU-Rx and consequently the average rate. In addition, having a greater capacity and a greater quantity of harvested energy reduces the likelihood that the agent will be penalized for choosing a transmission power that exceeds the energy content of the battery. {\color{black} Furthermore}, it can be seen that the performance diminishes as the TS factor $\rho$ rises. This is due to the fact that a greater $\rho$ indicates less time allowed for data transmission $(1-\rho)$ and, therefore, a lower achievable data rate (a lower $\mu$ in (\ref{rate1}) and (\ref{rate2})). However, the SU-Tx battery has sufficient energy to compensate for power losses and it has benefited from the interference of the PUs' transmissions as a source of energy through TS-EH.  This indicates that the SU must decide on the TS factor $\rho$ depending on whether it seeks to enhance energy efficiency or system reliability. 

\section{Conclusions}
This paper presents an underlay CRN in which a pair of SUs are assumed to interact through this access mode in the presence of two PU transmitters. The SU transmitter extracts energy from the RF signals emitted by the PU transmitters and from ambient RF sources. To optimize the average rate of SUs, a DQN strategy is utilized. Using this algorithm, the agent interacts with the surrounding environment and balances energy harvesting and data transmission. The suggested method outperforms a random strategy, and it converges after a certain number of episodes, according to our findings. Furthermore, since the SU-Rx is   affected by PU1's interference, our findings indicate that the reward increases with a reduction in the number of time slots established for PU1's broadcasts. Moreover, it is shown that when the battery capacity of the SU transmitter increases, system performance improves. In addition, the average rate improves when the transmission threshold set by the SU-Tx for determining the source of harvesting rises. It was also observed that the SU transmitter needs to decide on the value of the time switching factor depending on the objective of the system, i.e., improving the energy efficiency or system's reliability.  Lastly, given the interference constraints of PUs and limited battery capacity, the long-term achievable rate is maximized.




\end{document}